\documentclass[twocolumn,showpacs,preprintnumbers,amsmath,amssymb]{revtex4}
\usepackage{graphicx}
\usepackage{dcolumn}
\usepackage{bm}

\begin{document}

\preprint{APS/123-QED}
\title{Cold atoms in a high-$Q$ ring-cavity}

\author{D. Kruse}
\email{kruse@pit.physik.uni-tuebingen.de}
\author{M. Ruder}
\author{J. Benhelm}
\author{C. von Cube}
\author{C. Zimmermann}
\author{Ph. W. Courteille}
\affiliation{Physikalisches Institut, Eberhard-Karls-Universit\"at
T\"ubingen,
\\Auf der Morgenstelle 14, D-72076 T\"ubingen, Germany}

\author{Th. Els\"asser, B. Nagorny, and A. Hemmerich}
\affiliation{Institut f\"ur Laser-Physik, Universit\"at Hamburg,
\\Jungiusstr. 9, D-20355 Hamburg, Germany}

\date{\today}

\begin{abstract}
We report the confinement of large clouds of ultra-cold $^{85}Rb$
atoms in a standing-wave dipole trap formed by the two
counter-propagating modes of a high-$Q$ ring-cavity. Studying the
properties of this trap we demonstrate loading of higher-order
transverse cavity modes and excite recoil-induced resonances.
\end{abstract}

\pacs{32.80.Pj, 32.70.Jz, 42.50.Vk, 42.50.-p}

\maketitle

The tremendous progress of ultra-cold atomic physics in the past decades is
mainly due to the invention of powerful techniques for trapping and cooling
atoms. Besides magnetic traps, optical forces are used to store atoms in
tightly focused red-detuned laser beams, in the antinodes of a standing wave
formed by two counter-propagating laser beams or even in three-dimensional
optical lattices. Spontaneous scattering processes are avoided by tuning the
lasers far from resonance. High light intensities are then needed to keep the
atom-field coupling strong. Large intensity enhancements are achieved with
optical cavities \cite{Mosk01}. The coupled system made up of the cavity mode
and the atoms exhibits fascinating novel features
\cite{Ye99,Pinske00,Sinatra98}. For example, cooling procedures have been
proposed \cite{Mossberg91,Cirac95, Horak97,Vuletic00}, where the kinetic
energy of the atoms is dissipated via the decay of the cavity field to which
the atoms are coherently coupled, rather than via spontaneous decay of atomic
excitation energy. This bears the advantage that the cooling is only limited
by the technical parameter of the cavity decay rate. As the cooling procedure
is quite insensitive to the details of the atomic level structure, it should
work even for molecules.

In this paper, we study atomic trapping in a high-$Q$ ring-cavity with both
propagation directions pumped. This kind of cavity is particular in the
following sense. The optical dipole force exerted by a standing wave on atoms
can be understood in terms of a momentum transfer by coherent redistribution
(via Rayleigh scattering) of photons between the counter-propagating laser
beams. In multimode configurations, \emph{e.g.} a ring-cavity, the photon
redistribution can occur between \emph{different} modes \cite{Gangl00b}. This
implies that the atoms have a noticeable backaction on the optical fields and
that the atomic dynamics can be monitored as an intensity
\cite{Kozuma96,Raithel98} or a phase imbalance between the modes. In a
ring-cavity, the scattering of a photon between the counter-propagating modes
slightly shifts the phase of the standing wave. This shift, being strongly
enhanced by a long cavity life time, is sensed by all atoms trapped in the
standing wave. Consequently, the simultaneous interaction of the atoms with
the two field modes couples the motion of all atoms \cite{Hemmerich99,Gangl00}%
, so that in contrast to conventional standing wave dipole traps,
the atoms do not move independently of each other.

\begin{figure}
\vspace{7cm}
\includegraphics{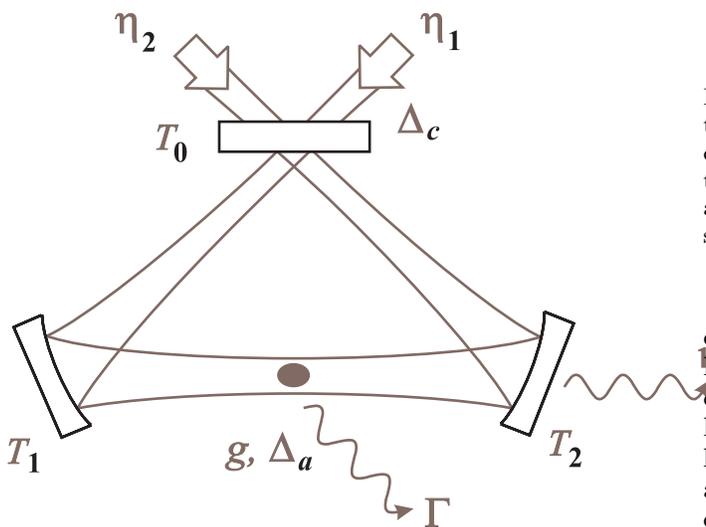}
\caption{Geometry of the ring-cavity. The atomic cloud is located
in the free-space waist of the cavity mode. The system is
characterized by the pumping parameter $\eta_{i}$, the atom-field
coupling $g$, the laser detuning with respect to the cavity
$\Delta_{c}$\ and to the atomic resonance $\Delta_{a}$, and the
decay width of
the atom $\Gamma$\ and of the cavity field $\kappa$.}%
\label{fig:1}
\end{figure}

In a recent article~\cite{Nagorny03} we have shown that a
far-detuned optical lattice can be formed inside a high-$Q$
ring-cavity and that heating due to intensity fluctuations can be
kept at very low levels despite of the need to maintain a sharp
resonance condition by a high-bandwidth servo control. Here we
present a setup where we show that the atomic motion can be probed
non-destructively and \emph{in-situ} by recoil-induced resonances
\cite{Courtois94}. There is an important reason to look for
methods to study the atomic motion which do not rely on sudden
changes of the intra-cavity intensity. This is because the long
lifetime of a high-$Q$ cavity inhibits the non-adiabatic
switch-off of deep cavity dipole traps and thus impedes a straight
forward interpretation of time-of-flight (TOF) images to determine
the axial temperature.

We fill our ring-cavity dipole trap with $^{85}$Rb atoms from a
standard magneto-optical trap (MOT) which is loaded from a vapor
generated by a rubidium dispenser. Typically, we load $10^{8}$
atoms into the MOT at temperatures around $140~\mu$K with a vapor
pressure of $3\times10^{-9}$~mbar.

\begin{figure}[h]
\centerline{\scalebox{0.6}{\includegraphics{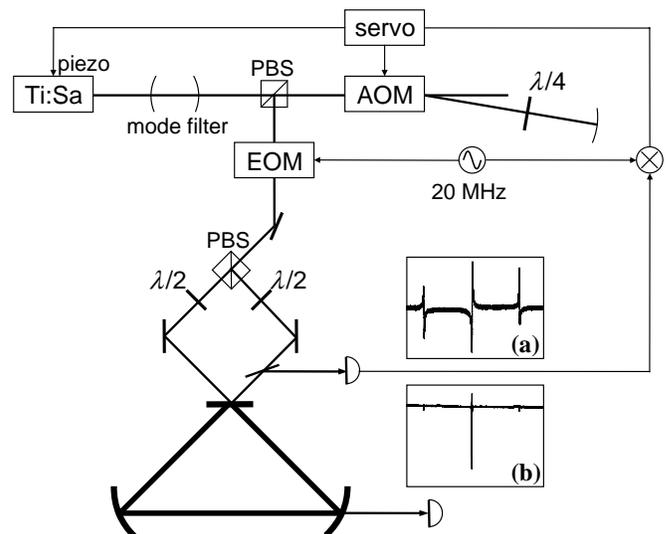}}}%
\caption{Experimental setup for pumping both directions of the ring-cavity
while locking the laser to a resonator eigenfrequency via the
Pound-Drever-Hall technique. The AOM and the piezo rule out the laser
frequency fluctuations. Also shown are the demodulated reflection signal (a)
and the transmission signal (b) of the cavity.}%
\end{figure}

The geometry of the ring cavity is shown in Fig.~1. The
transmissions $T_{i}$ of the mirrors depend on the (linear)
polarization of the light modes. For $s $-polarization
$T_{0}=27\times10^{-6}$, $T_{1}=T_{2}=2\times10^{-6}$, while for
$p$-polarization $T_{0}=2200\times10^{-6}$,
$T_{1}=T_{2}=9\times10^{-6}$. The experiments presented here are
carried out with $p$-polarized light and a measured finesse of
$F=2500$, which corresponds to an intracavity intensity decay rate
of $2\pi\times 1.4~$MHz. However, by simply rotating the
polarization of the light injected into the cavity, we can switch
to $s$-polarization, where we measure using the ring-down method a
much higher finesse of $F=170000$ corresponding to $2\pi\times
21~$kHz. The round-trip length of the ring-cavity is $L=85$~mm,
the beam waists in horizontal and in vertical direction at the
location of the MOT are $w_{v}=129~\mu$m and $w_{h}=124~\mu$m,
respectively. This yields a cavity mode-volume of
$V_{mode}=\frac{\pi}{4}Lw_{v}w_{h}=1$~mm$^{3}$. The intracavity
power $P$, largely enhanced by the factor $F/\pi$ to values around
$10~$W, gives rise to an optical potential with a depth of
$k_{B}\times1.4$~mK at the wavelength $799~ $nm. The radial and
axial secular frequencies in the harmonic region close to the
center of the trap are $\omega_{rad}/2\pi=640$~Hz and
$\omega_{ax}/2\pi=450$~kHz.

The ring-cavity is driven by a titanium-sapphire (Ti:Sa) laser delivering up
to $2~$W output power into three optical modes separated by $1.2~$GHz
\cite{Zimmermann95}. The central mode is filtered by an external confocal
Fabry-Perot etalon. The Ti:Sa laser is locked to one of the eigenfrequencies
of the ring cavity using the Pound-Drever-Hall locking technique
\cite{Pound46,Drever83} (see Fig.~2). A feedback servo drives a piezo-electric
transducer mounted in the Ti:Sa laser cavity, whose frequency response is
limited to $10$ kHz. Faster fluctuations of the laser frequency are balanced
by means of an external double-passed acousto-optic modulator (AOM). The servo
bandwidth of $1$~MHz is limited by a $100~$ns time delay in the response of
the AOM. The laser frequency is stable enough to yield intensity fluctuations
observed in the cavity transmission signal below $2~\%$ even in the high
finesse case.

The dipole trap is permanently operated, because keeping the Ti:Sa
laser locked requires a certain amount of light inside the cavity.
The standing wave dipole trap is loaded from a spatially
overlapping MOT for a period of $15$~s. Before switching off the
MOT, we apply a $40$~ms temporal dark MOT stage \cite{Ketterle93}
by increasing the detuning of the MOT laser beams to $-90~$MHz and
reducing the intensity of the repumping beams. For the conditions
given above we typically capture $3\times10^{7}$ atoms distributed
over $10000$ antinodes of the standing wave. The temperature of
the atomic ensemble is measured using the TOF method. We suddenly
switch off the dipole trap and image the shadow that the cloud
imprints on a weak probe beam after a period of ballistic
expansion. During the expansion, the initial momentum distribution
of the cloud evolves into a density distribution, whose radial
width yields the temperature of the cloud. Depending on the
potential well depth, we obtain temperatures between $70$ and
$280$~$\mu$K, corresponding to roughly $1/5$ of the well depth.
The peak density is typically $3\times10^{12}$~cm$^{-3}$. The
lifetime of the dipole trap measured at $799~$nm is $0.5$~s. A
thorough investigation of trap loss processes is presented in a
similar system in Ref.~\cite{Nagorny03}.

With a slight misalignment of the incoupled laser beam, the
ring-cavity can be locked to higher-order transverse modes, into
which the atoms can settle. Fig.~3 shows absorption pictures of
atomic clouds confined in different higher-order transverse modes.
Such modes exhibit an enhanced surface-to-volume ratio, which may
prove advantageous for forced evaporation.

\begin{figure}[h]
\centerline{\scalebox{0.63}{\includegraphics{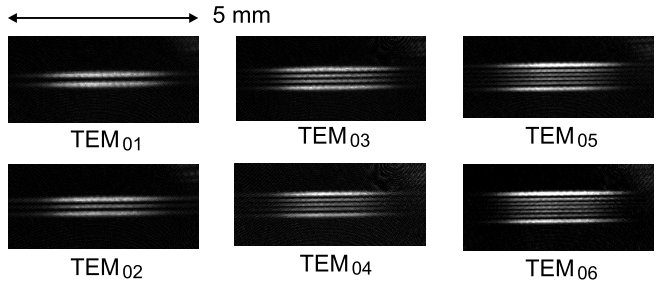}}}%
\caption{Absorption pictures of atoms stored in higher-order transverse cavity
modes.}%
\end{figure}

We perform spectroscopy of recoil-induced resonances (RIR)
\cite{Guo92,Courtois94,Meacher94}, \emph{i.e.} we probe 2-photon Raman
transitions between two velocity states of the same atom. Two Raman beams
$\mathbf{k}_{1}$ and $\mathbf{k}_{2}$ enclosing a small angle $\theta
=13.1^{\circ}$ are radiated onto the atomic cloud such that the difference
vector $\mathbf{q=k}_{1}-\mathbf{k}_{2}$ is oriented nearly parallel to the
dipole trap symmetry axis $\hat{z}$. The two beams, whose frequencies are
$\omega_{1}$ and $\omega_{2}$, give rise to a standing wave with periodicity
$2\pi/q$ slowly moving in $\hat{z}$ direction with velocity $v_{z}%
=\Delta\omega/q$, where $q=(k_{1}+k_{2})\sin(\theta/2)$ and
$\Delta \omega=\omega_{1}-\omega_{2}$. The light wave leads to a
periodic dipole potential for the atoms which in our experiment
has a well depth of $60$~$\mu $K in units of $k_{B}$. Only atoms
satisfying the energy and momentum conservation requirement can
undergo Raman transitions, \emph{i.e.} only atoms moving
synchronously to the standing wave can scatter light from one beam
into the other. This scattering is monitored as an intensity
variation in one of the beams. The net rate for scattering from
beam 1 into beam 2 may be written as
$W(v_{z}=\Delta\omega/q)=\hbar\pi/2\cdot\Omega_{R}^{2}\cdot N\cdot
\partial\Pi/\partial v_{z}$, where $\Pi(v_{z})$ denotes the Maxwell-Boltzmann
momentum distribution, $N$ the number of atoms and
$\Omega_{R}=\Omega _{1}\Omega_{2}/2\Delta$ with the resonant
Rabi-frequencies $\Omega_{i}$ for each Raman beam. By varying
$\Delta\omega$ the derivative of the Maxwell-Boltzmann momentum
distribution is scanned from which the temperature can be derived
\cite{Meacher94}. Trace (a) of Fig.~4 shows such RIR-scans
recorded on a cloud $200$~$\mu$s after being released from the
dipole trap. Trace (b) has been calculated assuming a $100$~$\mu$K
cold cloud.

The scan rate must be judiciously chosen \cite{Fischer01}. If the
scan rate is too slow, the atoms are notably redistributed between
the velocity classes while scanning. The above expression shows
that the scattering process preferentially occurs towards higher
velocities, so that although the momentum transfer is quite small,
the cloud is slightly heated under the influence of the Raman
beams. If, on the other hand, the scan rate is too fast the signal
can be strongly distorted and a ringing type oscillation is
observed \cite{Note}. For an untrapped cloud of atoms first
indication of ringing can be observed for scan rates above $2.1$
kHz/$\mu$s. For trapped atoms, however, ringing is already very
pronounced at this rate (Fig.~4c). Ringing is a general feature of
resonant phenomena and can be observed for a simple mechanical
oscillator as well as in an optical resonator or for standard two
level quantum systems. The critical scan rate for ringing is
roughly given by the square of the linewidth. For slower scan
rates the system can follow adiabatically, \emph{i.e.} the sweep
time across the resonance is longer than the inverse linewidth.

\begin{figure}[h]
\centerline{\scalebox{0.47}{\includegraphics{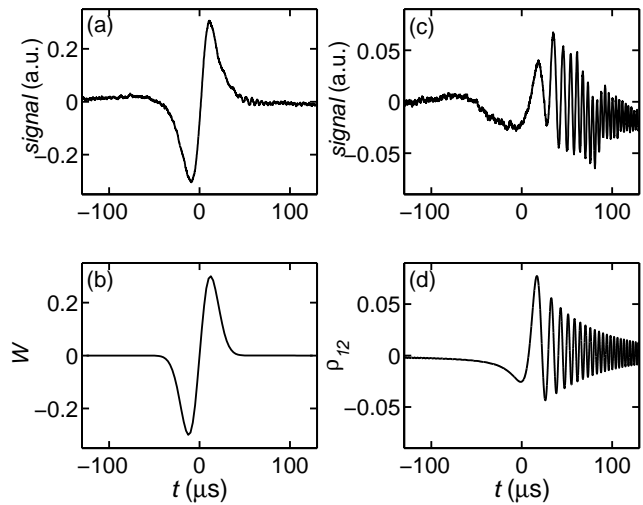}}}\caption{Trace
(a): Spectrum of RIR resonances recorded on \emph{free} atoms. The
intensity of Raman beam with $\omega_{2}$ is measured, while the
frequency of beam with $\omega_{1}$ is scanned. Both beams were
detuned by $-110$~MHz from resonance and their peak intensities
were $50~$mW/cm$^{2}$. The scan rate was $2.1$~kHz/$\mu$s. Trace
(b): Calculated transition rate for the same parameters assuming a
$100~\mu$K cold cloud. Trace (c): Same conditions but
recorded on \emph{trapped} atoms. The trap had a well-depth of $U_{0}%
=h\times30$~MHz~$=k_{B}\times1.6~$mK corresponding to secular
frequencies $\omega_{z}=2\pi\times700$~kHz and
$\omega_{r}=2\pi\times1$~kHz. Trace (d) shows a simulation (see
text) of the susceptibility of a two-level atom subject to a laser
quickly swept over its resonance. The chosen decay rate was
$\Gamma=2\pi\times5$~kHz and the Rabi-frequency $\Omega=0.1~\Gamma$.}%
\end{figure}

In our experiment we observe that the critical scan rate is
dramatically reduced if the atoms are trapped in the standing wave
dipole trap. A qualitative explanation of this effect is possible
by using the following simple picture. For an untrapped cloud of
atoms the potential generated by the Raman beams has its principle
effect on the atoms near zero detuning. Then, the speed of the
potential wave is sufficiently slow for the atoms to react and
form a standing wave like density pattern that mirrors the shape
of the potential wave. For positive detunings the potential wave
starts moving with increasing speed relative to the static atomic
distribution and the atoms within the density modulation are
subjected to an oscillating dipole force. This force is
accompanied by a periodic photon redistribution between the Raman
beams which is observed in the experiment. The ringing fades away
as the atomic density modulation slowly decays due to thermal
drifts of the atoms. For trapped atoms this picture does not seem
to apply since the atoms are strongly confined within the
antinodes of the resonator dipole trap such that a density pattern
can not form. However, if we neglect the weak confinement in
radial direction and thus regard a thermal distribution of atoms
that can still propagate freely along the valleys of the resonator
dipole trap, this is only true if $\mathbf{q}$ is strictly
parallel to the symmetry axis of the dipole trap. With only a
slight misalignment of $\mathbf{q}$ by an angle $\phi$ the atoms
can follow the force of the Raman potential by travelling along
those valleys. In order to form a density modulation similar to
the untrapped case the atoms now have to propagate a distance that
is longer by a factor $f=1/\sin\phi$. Similarly, the lifetime of
the modulation pattern is enhanced by the same factor since the
atoms have to drift over a longer distance with the same thermal
velocity. The critical scan rate for ringing is therefore reduced
by the factor $f^{2}$. The same situation can be explained from a
different perspective. Since the phase speed of the Raman
potential along the valleys is enhanced by $f$ the atoms that
interact with the Raman potential have to move with a velocity
given by $f\cdot\Delta\omega/q$. The resulting reduction of the
linewidth by $1/f$ can easily be understood by the fact that for
an axially confined atom undergoing a 2-photon Raman transition
only the radial momentum component with $q_{r}=q/f$ has to satisfy
the momentum conservation requirement. From the data we can
estimate the coherence lifetime to be $100~\mu$s. This compares
well to the above described thermal decay of the density
distribution if we assume a misalignment angle of
$\phi=3^{\circ}$.

A full quantitative description is beyond the scope of this paper,
however it is interesting to note that the data can be well
described by the dynamics of a degenerate two level quantum
system. The lower curve in Fig.~4(d) shows the coherence
$\operatorname{Im}\rho_{12}$ calculated by numerical integration
of two-level Bloch-equations ($\rho$ is the density matrix). The
resonance occurs at $\Delta\omega=0$. At the beginning of the
scan, where $\Delta\omega$ is far from resonance, the population
of the excited level $\rho_{22}$ is just too small and the system
does not react to the light field. As soon as $\Delta\omega$
passes through the resonance, the coherence $\rho_{12}$ is excited
and can now be driven by the laser even when $\Delta\omega$ is
tuned far away. The two levels can be identified with two velocity
states that are coupled by the Raman-beams. In the experiment, the
thermal distribution causes an inhomogeneous broadening that is
accounted for in the simple two level model by introducing an
effective decay rate $\Gamma$.

To conclude we introduced our ring-cavity standing wave dipole trap and
characterized it by lifetime and by temperature measurements. We have shown
that by driving recoil-induced resonances, we can excite and probe the motion
of atoms trapped in the ring-cavity field. In the near future we plan to look
for the expected feedback of the atomic motion on the standing light wave and
for interatomic coupling induced by the standing wave. It is also tempting to
explore the predicted new types of cavity cooling in view of their aptitude of
cooling below the threshold of quantum degeneracy. Bose-Einstein condensates
are very appealing objects in the context of ring-cavity studies. \emph{E.g.}
Meystre and coworkers \cite{Moore99} have discussed the use of ring-cavities
for recycling superradiant light pulses produced by Rayleigh-scattering off
condensates \cite{Inouye99} and predict for such systems the possibility of
mutual coherent quantum control between optical and matter-wave modes.

We acknowledge financial support from the Landesstiftung
Baden-W\"urttemberg and the Deutsche Forschungsgemeinschaft.


\begin{thebibliography}{99}
\bibitem{Note}Similar observations have been made in the case of untrapped
atoms for a very rapid change of the detuning $\Delta\omega$ from zero to a
non-zero value \cite{Kozuma95}. In this case, the observed oscillations have
been explained with the development of a stationary atomic density grating
periodically sensed by the Raman standing wave moving with velocity
$\Delta\omega$.
\end{thebibliography}
\end{document}